# SIDE-real: Supernova Ia Dust Extinction with truncated marginal neural ratio estimation applied to real data

Konstantin Karchev[1]★, Matthew Grayling[2], Benjamin M. Boyd,[2] Roberto Trotta[1,3,4,5]★, Kaisey S. Mandel[2,6] and Christoph Weniger[7]

[1]*Theoretical and Scientific Data Science, Scuola Internazionale Superiore di Studi Avanzati (SISSA), Via Bonomea 265, I-34136 Trieste, Italy*
[2]*Institute of Astronomy and Kavli Institute for Cosmology, Madingley Road, Cambridge CB3 0HA, United Kingdom*
[3]*Astrophysics Group, Physics Department, Blackett Lab, Imperial College London, Prince Consort Road, London SW7 2AZ, United Kingdom*
[4]*INFN – National Institute for Nuclear Physics, Via Valerio 2, I-34127 Trieste, Italy*
[5]*Italian Research Center on High-Performance Computing, Big Data and Quantum Computing, Via Magnanelli 2, I-40033 Casalecchio di Reno (BO), Italy*
[6]*Statistical Laboratory, DPMMS, University of Cambridge, Wilberforce Road, Cambridge CB3 0WB, UK*
[7]*Gravitational and Astroparticle Physics Amsterdam (GRAPPA), University of Amsterdam, Science Park 904, NL-1098 XH Amsterdam, the Netherlands*

## ABSTRACT

We present the first fully simulation-based hierarchical analysis of the light curves of a population of low-redshift type Ia supernovæ (SNæ Ia). Our hardware-accelerated forward model, released in the Python package slicsim, includes stochastic variations of each SN's spectral flux distribution (based on the pre-trained BayeSN model), extinction from dust in the host and in the Milky Way, redshift, and realistic instrumental noise. By utilizing truncated marginal neural ratio estimation (TMNRE), a neural network-enabled simulation-based inference technique, we implicitly marginalize over 4000 latent variables (for a set of $\approx 100$ SNæ Ia) to efficiently infer SN Ia absolute magnitudes and host-galaxy dust properties at the population level while also constraining the parameters of individual objects. Amortization of the inference procedure allows us to obtain coverage guarantees for our results through Bayesian validation and frequentist calibration. Furthermore, we show a detailed comparison to full likelihood-based inference, implemented through Hamiltonian Monte Carlo, on simulated data and then apply TMNRE to the light curves of 86 SNæ Ia from the Carnegie Supernova Project, deriving marginal posteriors in excellent agreement with previous work. Given its ability to accommodate arbitrarily complex extensions to the forward model, e.g. different populations based on host properties, redshift evolution, complicated photometric redshift estimates, selection effects, and non-Ia contamination, without significant modifications to the inference procedure, TMNRE has the potential to become the tool of choice for cosmological parameter inference from future, large SN Ia samples.

## 1 INTRODUCTION

Type Ia supernovæ (SNæ Ia) are extremely bright and remarkably consistent – after standardization – stellar explosions, which enabled the discovery of the Universe's accelerated expansion (Riess et al. 1998; Perlmutter et al. 1999) and have since been the main probe into the properties – and ultimately, nature – of dark energy (Huterer & Shafer 2018). They also form a vital rung in the cosmic distance ladder, allowing a local measurement of the Hubble constant (Riess et al. 2022), which is famously in tension with estimates based on observations of the cosmic microwave background (CMB; Di Valentino et al. 2021).

The process of standardization – crucial for using SNæ Ia as distance indicators – involves correlating observed properties of a supernova, such as its colour and light-curve shape, to its intrinsic luminosity, so that it can be compared to the SN's observed brightness (Phillips 1993; Tripp 1997, 1998). However, the latter is also affected by the amount and properties of interstellar dust that surrounds the supernova and diminishes its brightness. Crucially, the amount of dust extinction depends on wavelength, impacting both the brightness and colour of the SN. Historically, SN Ia models like the de-facto standard SALT (Guy et al. 2005, 2007; Betoule et al. 2014; Kenworthy et al. 2021; Taylor et al. 2021) made no distinction between the intrinsic colour of a SN and the reddening induced by dust and instead explained the twofold effect of extinction by correlating the SN's luminosity with the single parameter measuring its apparent colour (*bluer–brighter*).

Disentangling dust-induced reddening and dimming from the pure dimming effect of distance on observed SN Ia brightnesses can lead to better standardization (and thence, to more powerful cosmological constraints), as illustrated by Mandel et al. (2017), Brout & Scolnic

★ E-mail: kkarchev@sissa.it (KK); r.trotta@imperial.ac.uk (RT)

(2021) via post-factum corrections of SALT parameters. Brout & Riess (2023) recently reviewed how the issues surrounding dust are tackled in conventional SN Ia analyses.

On the other hand, detailed dust modelling is a principal component of BayeSN (Mandel et al. 2009; Mandel, Narayan & Kirshner 2011; Thorp et al. 2021; Mandel et al. 2022; Ward et al. 2023a), a probabilistic SN Ia model that furthermore captures the *broader–brighter* correlation observed in SNæ Ia into a trainable spectro-temporal flux distribution, thus obviating the need for Tripp corrections involving the magnitude, stretch, and dust parameters (Phillips 1993; Tripp 1997, 1998). BayeSN also exploits observations in the near-infrared (IR), where extinction is reduced and only weakly dependent on the dust properties, as a convenient 'anchor' for measuring the SNæ's brightness (Avelino et al. 2019).

Astrophysical properties of galaxies are also thought to influence the *population* of SNæ Ia they host. Correlations between (standardized) SN Ia absolute magnitudes and colours on one side and the host overall stellar population age, star-formation rate, metallicity, and stellar mass (Kelly et al. 2010; Sullivan et al. 2010; Childress et al. 2013; Chung et al. 2023), as well as the SN's location within the host and the local host properties (e.g. Rigault et al. 2013, 2015, 2020; Jones, Riess & Scolnic 2015; Moreno-Raya et al. 2016b, a; Hill et al. 2018; Jones et al. 2018; Kim et al. 2018; Roman et al. 2018; Kim, Kang & Lee 2019; Rose, Garnavich & Berg 2019; Kelsey et al. 2021) on the other side have been empirically observed, although a definitive explanation of the causal channels is still outstanding.

Studying populations of SNæ Ia has often involved a two-step process whereby first, the properties of individual objects are inferred, and then their distributions are examined. Modern analyses methodologies like BAHAMAS (March et al. 2011; Shariff et al. 2016), UNITY (Rubin et al. 2015, 2023), Ma, Corasaniti & Bassett (2016), Steve (Hinton et al. 2019), and BayeSN are instead *hierarchical*: they include parameters describing the SN Ia population and infer them simultaneously with those of the individual SNæ. While the former three use a few highly compressed summary statistics instead of the full light curves, BayeSN models probabilistically the full spectro-temporal flux distribution of SNæ Ia. A middle ground is BIRD-SNACK (Ward et al. 2023b), which models hierarchically the light curves around peak with a restricted set of colour-related parameters. The impact of Bayesian hierarchical modelling (BHM) of dust extinction on inferred SN Ia distance estimates (used subsequently for cosmology) was demonstrated by Mandel et al. (2017), Thorp & Mandel (2022, hereafter TM22). Recently, Grayling et al. (2024) extended the analysis to inferring SN Ia models separately for high- and low-mass hosts, while Thorp et al. (2024) considered an evolution of dust properties by comparing a low- to a higher-redshift sample.

Future surveys, like those performed by the upcoming Roman Space Telescope (WFIRST; Hounsell et al. 2018) and Vera Rubin Observatory (LSST; LSST Science Collaboration 2009; Ivezić et al. 2019), are expected to detect hundreds of thousands of supernovæ (Ia and non-Ia), greatly reducing purely statistical uncertainties and highlighting the need for principled treatment of systematic effects that avoids approximations and ad hoc assumptions. Apart from requiring scalable computational methods for their analysis, the orders-of-magnitude increase in the number of SN Ia candidates will also introduce new modelling challenges, mainly related to the inability to have spectroscopic follow-up for any but a small fraction of the detected transients. Among those challenges are complicated (non-Gaussian, multimodal) posterior estimates for the objects' redshifts, contamination of the sample due to mis-classification (Kunz, Bassett & Hlozek 2007), covariate shift (Moreno-Torres et al. 2012, see also Revsbech, Trotta & van Dyk 2018; Autenrieth et al. 2023) between the full sample and a high-quality/spectroscopic sample used to train SN Ia models, and selection effects like Malmquist bias (Malmquist 1922, 1925), whereby the preference for detecting brighter objects skews and shifts the distribution of properties of detected SNæ Ia away from that of the whole population.

In principle, these effects can be included in likelihood-based analyses, with two approaches prevailing. Some studies – see e.g. the recent photometry-only cosmological analysis of Popovic et al. (2024) – rely on simulations to derive various de-biasing and correction factors (Kessler & Scolnic 2017), which, however, are at best correct and de-biasing *on average*. Furthermore, the focus of this procedure on observed SN properties – inherited from the tradition of standardization – makes it difficult to extend to properties of the hosts (see e.g. Popovic et al. 2021, 2023). On the other hand, hierarchical likelihood-based analyses (Rubin et al. 2015; Hinton et al. 2019; Rubin et al. 2023, see also March et al. 2018) can only handle selection effects by transferring them to the level of the unobserved latent parameters, which necessitates a plethora of 'counter-intuitive' (Rubin et al. 2015) assumptions and approximations to the selection likelihood, which is intractable for realistic selection criteria applied to light curves.

The framework of simulation-based inference (SBI) is a powerful alternative to likelihood-based methods which allows simulations to be used to directly derive quantities of interest: for example, marginal posteriors for the cosmological parameters or those describing the dust population. Stemming from approximate Bayesian computation (ABC; for a review, see e.g. Sisson, Fan & Beaumont 2018), SBI comprises a rapidly expanding collection of techniques centred around the idea of representing the data-generating process not through numerical evaluation of the likelihood but rather through example (mock) data stochastically simulated with known parameters.

Because the pairs of parameters and simulated data implicitly encode the full likelihood, it is possible to define arbitrary parameter sub-spaces in which to perform marginal simulation-based inference. This is especially beneficial when analysing large collections of observed objects, for which likelihood-based methods require exploring a large latent parameter space even if one is only interested in a few global parameters: in the case of SN Ia cosmology, one needs to infer, e.g. the individual redshifts, stretches, colours, etc. of all SNæ, only to eventually marginalize them out and obtain a two-dimensional marginal posterior for the cosmological parameters of interest. While all latent parameters still need to be stochastically sampled in the context of SBI, this only adds a linear complexity of simulating all objects, instead of the at least quadratic – but often worse: see e.g. Handley, Hobson & Lasenby (2015, fig. 5) – scaling required to map out high-dimensional spaces.

SBI also allows seamless inclusion of many aspects of the model for which a numerical likelihood is intractable or computationally impractical: for example, contamination and selection bias can be represented through contaminated and biased mock catalogues, created by performing on simulated data the same classification procedure and enforcing the same selection criteria as on real observations. Importantly, because the true input parameters to the simulations are known, these catalogues can be used to derive accurate (de-biased and de-contaminated) posteriors.

Different SBI flavours use different procedures to convert parameter–simulated data pairs into posteriors conditioned on the observed data. ABC, for example, accepts or rejects input parameters based on the similarity of simulated and observed data, but this is computationally wasteful and requires bespoke distance mea-sures for all but the simplest types of data. More recent methods employ neural networks (NNs) to interpolate in data space and have thus proven much more efficient in terms of simulation budget. The inference network can be trained to either emulate the likelihood: neural likelihood estimation (NLE), approximate the posterior of interest: neural posterior estimation (NPE), or estimate the likelihood-to-evidence ratio: neural ratio estimation (NRE).[1] Whereas NLE and NPE require density estimation, NRE transforms Bayesian inference into a classification task, thus allowing the greatest freedom to the neural network architecture. We use a sequential modification known as truncated marginal neural ratio estimation (TMNRE; Miller et al. 2021), which further optimizes simulation use and network expressivity by focusing on regions of parameter space consistent with the observed data, while trivially composing with marginalization and preserving local amortization (see below).

Neural SBI techniques are also amortized: i.e. rather than the particular posterior given the observed data, they learn how to derive the posterior from any data. This, in combination with a nearly instantaneous evaluation speed once trained, allows neural SBI to be validated on large sets of simulations, either with random parameters from the priors (Hermans et al. 2022) or at fixed parameter values, i.e. in a frequentist context. Amortization can also be used during training to ensure proper Bayesian coverage properties (Delaunoy et al. 2022).

Carried by the rapid developments in the field of deep learning, neural SBI has recently grown in popularity, with prominent applications in cosmology, particle physics, and beyond. In particular, TMNRE has been applied to gravitational waves (Alvey et al. 2023b, 2024; Bhardwaj et al. 2023), 21-cm cosmology (Saxena et al. 2023), stellar streams (Alvey, Gerdes & Weniger 2023a), point-source population studies (Anau Montel & Weniger 2022), and strong lensing (Anau Montel et al. 2022; Coogan et al. 2024). On the other hand, examples of SBI from the field of SN Ia analysis include early uses of ABC (Weyant, Schafer & Wood-Vasey 2013; Jennings, Wolf & Sako 2016), ABC with evolutionary optimization (Bernardo et al. 2023), and, more recently, of NPE (Alsing et al. 2019; Alsing & Wandelt 2019; Villar 2022; Wang et al. 2022, 2023; Chen et al. 2023), and TMNRE (Karchev, Trotta & Weniger 2023a, hereafter SICRET).

The analysis in SICRET (and the majority of other works mentioned above[2]) is based on pre-derived summary statistics (e.g. from SALT) and focused on scaling it to the size of near-future surveys ($\sim 10^5$ SNæ Ia). The present work, in complement, extends the methodology in terms of data and modelling complexity by training a neural network to summarize raw light-curve data in a way that is optimal for the particular inference task, thus circumventing the expensive fitting stage present in all current studies. While our focus is on marginal population-level inference (e.g. of global dust properties and, eventually, cosmology), TMNRE also allows simultaneous inference of all object-level (local) parameters in the BHM: the very ones that serve as summary statistics in downstream tasks. Finally, explicitly modelling the SN light curves is indispensable if one wants to account for selection effects (introduced by criteria defined in terms of raw observations) and non-Ia contamination. In a forthcoming work, we plan to address these two issues, the final hurdle to achieving fully simulation-based SN Ia cosmology.

We describe our forward simulator of SN Ia light curves in Section 2 and briefly introduce TMNRE in Section 3, detailing the network we use in Section 3.1. In Section 4, we validate the results from our method against those obtained via Hamiltonian Monte Carlo (HMC) with simulated data that mimics the Carnegie Supernova Project (CSP) survey before presenting results on the real SN Ia light curves in Section 5 and concluding in Section 6.

## 2 slicsim: REALISTIC SN IA LIGHT CURVE SIMULATIONS FOR MACHINE LEARNING

SBI analyses are empowered by the realism of the simulations they employ. This work improves in this respect compared to SICRET by utilizing a simulator that generates realistic SN Ia light curves while incorporating uncertainties from three different levels: observational noise, the hierarchy of stochastic parameters, and the residual stochasticity of the SN's spectro-temporal flux surface.

Alongside the development of SN Ia light-curve models to be used in likelihood-based analyses (e.g. SALT(Guy et al. 2005, 2007; Betoule et al. 2014; Kenworthy et al. 2021; Taylor et al. 2021), SNEMO (Saunders et al. 2018), BayeSN (Mandel et al. 2009; Mandel et al. 2011; Thorp et al. 2021; Mandel et al. 2022; Ward et al. 2023a), SUPAERNOVA Stein et al. 2022), a number of general frameworks for realizing forward simulations using any given model, notably SNANA (Kessler et al. 2009) and sncosmo (Barbary et al. 2016), have emerged.

In this work, we present a new light-curve simulation framework, slicsim,[3] built from the ground up for close interoperability with modern machine-learning frameworks. Based on PyTorch (Paszke et al. 2019), our simulator is trivially deployable on hardware accelerators like graphics processing units (GPUs), parallelizable, and automatically differentiable.[4]

A SN Ia light-curve simulation involves three main components: a source model, a number of propagation effects, and an instrument model. Its input is a 'survey specification': the time, passband, and a description of the instrument and observing conditions (see Section 2.5) of each pointing comprising the analysed data. The survey specification is kept constant, so that the simulator is implicitly conditioned on it, producing an ordered list of simulated flux measurements $\left[ [d^{s,i}]_{i=1}^{N^s_{\mathrm{obs}}} \right]_{s=1}^{N_{\mathrm{SN}}}$. This conditioning allows us to use the efficient inference network described in Section 3.1.

### 2.1 Intrinsic SN Ia spectral time-series: BayeSN

The simulation starts with the spectral flux distribution (total emitted energy per unit time and unit wavelength interval), $\Phi_{t,\lambda}$, of a supernova $s \in \{1, \ldots, N_{\mathrm{SN}}\}$, i.e. its brightness at each point in time and at each wavelength in the SN rest frame. We use the BayeSN model (Mandel et al. 2009; Mandel et al. 2011; Thorp et al. 2021; Mandel et al. 2022; Ward et al. 2023a), which decomposes the magnitude difference from the Hsiao et al. (2007) SN Ia template, $\Phi_{\text{Hsiao}}(t, \lambda)$, as:

$$-2.5 \log_{10}[\Phi^s(t, \lambda)/\Phi_{\text{Hsiao}}(t, \lambda)]$$
$$= -19.5 + \delta M^s + W_0(t, \lambda) + \theta_1^s W_1(t, \lambda) + \epsilon^s(t, \lambda), \quad (1)$$

with $t, \lambda$ in the SN frame. Note that while $W_0$ and $W_1$ are shared among all SNæ Ia, the residual perturbations $\epsilon^s$ are not, and this sets BayeSN apart from other models based on functional principal component analysis (PCA) like SALT and SNEMO.

---

[1]See Cranmer, Brehmer & Louppe (2020), Lueckmann et al. (2021) for overviews of the methods and references for each and https://simulation-based-inference.org/ and https://github.com/smsharma/awesome-neural-sbi for references to applications. We list relevant ones below.

[2]except Jennings et al. (2016), who derived a distance measure between sets of light curves that avoids the curse of dimensionality, and Villar (2022), who resorted to Gaussian process (GP) interpolation to analyse light curves using neural networks

[3]https://github.com/kosiokarchev/slicsim

[4]A trait we do not exploit in this work but which underpins modern inference methods like variational inference (see Karchev 2023 for a simple demonstration on a toy SN Ia problem), automatic differentiation of the model likelihood can also be used in conjunction with likelihood-free methods to derive optimal summaries (Charnock, Lavaux & Wandelt 2018) or aid in training (Brehmer et al. 2020; Zeghal et al. 2022).

The principal components $W_0(t, \lambda)$ and $W_1(t, \lambda)$ and the residual perturbation surface $\epsilon^s(t, \lambda)$ are defined via two-dimensional spline interpolation over a fixed grid $[\mathbf{t}_g, \boldsymbol{\lambda}_g]$ in time and wavelength. The spline knots $\mathbf{W}_k$ and $\mathbf{e}^s$ can be learnt from data after setting suitable priors. Here, we use the pre-trained BayeSN model from Mandel et al. (2022, hereafter M20), based on 79 nearby SNæ Ia with high-quality optical and near-IR observations. It defines a $6 \times 9$ grid spanning the ranges $t \in [-10; 40]$ d and $\lambda \in [0.3; 1.85]$ μm. We fix the $\mathbf{W}_k$ and the common covariance matrix $\boldsymbol{\Sigma}_e$ of the $\mathbf{e}^s$ to the M20 posterior means:

$$W_k(t, \lambda) = \text{Spline2d}(t, \lambda; \mathbf{t}_g, \boldsymbol{\lambda}_g, \mathbf{W}_k), \quad \mathbf{W}_k = \text{fixed}, k = 0, 1; \quad (2)$$

$$\epsilon^s(t, \lambda) = \text{Spline2d}(t, \lambda; \mathbf{t}_g, \boldsymbol{\lambda}_g, \mathbf{e}^s), \quad \mathbf{e}^s \sim \mathcal{N}(\mathbf{0}, \boldsymbol{\Sigma}_e). \quad (3)$$

Thus, the 44 free parameters controlling the *intrinsic* brightness of each SN Ia are $\delta M^s$, $\theta_1^s$, and the 42-component array $\mathbf{e}^s$ (the perturbations are fixed to zero at the extreme wavelengths of the grid, reducing the trainable parameters to $6 \times 7$).

### 2.2 Propagation effects: dust extinction

In the context of BayeSN, dust extinction from the SN host is modelled with the Fitzpatrick (1999, hereafter F99) law:

$$\Phi^s(t, \lambda) \rightarrow \Phi^s(t, \lambda) \times \left[\text{F99}(\lambda; R_V^s)\right]^{A_V^s}, \quad (4)$$

where $t, \lambda$ are still in the SN frame, $R_V^s$ is the F99 dust-law parameter, and $A_V^s$ is the optical depth of host-galaxy dust for SN $s$.

### 2.3 Propagation effects: redshift and distance

The wavelength of light from a supernova is affected by the relative motion of the source and observer, which is made up of: the motion of the supernova within its galaxy (which is often neglected), the peculiar velocity of the host galaxy and of the Milky Way with respect to the CMB, and the Sun and the Earth's own peculiar motions. Additionally, distant objects are affected by the expansion of the Universe and exhibit a *cosmological* redshift.

The effect of the *total* redshift[5] $z^s$ is simple: it shifts the spectral flux distribution in both wavelength ($\lambda \rightarrow \lambda_o = (1 + z^s) \times \lambda$) and phase ($t \rightarrow t_o = (1 + z^s) \times t$) and suppresses its intensity (threefold: once because of the redshift of photons to lower energies, once because $\Phi$ is a rate (and time is dilated), and once because spectral intervals also get dilated):

$$\Phi_o^s(t_o, \lambda_o) = \frac{\Phi^s[t_o/(1 + z^s), \lambda_o/(1 + z^s)]}{(1 + z^s)^3}. \quad (5)$$

---

[5]See e.g. Davis et al. (2011) for the formulæ for combining and/or disentangling the various redshifts.

In this work, we will assume to have perfect estimates $\hat{z}^s = z^s$ of the total redshifts, as appropriate for spectroscopically observed supernovæ (uncertainties in this case are on the order of $\sigma_z \approx 10^{-5}$). Most supernovae from future surveys, however, will only have redshift estimates from photometric observations of their host galaxy (thus, not including the SN's peculiar motion), which are highly uncertain and expected to deviate significantly from the Gaussian approximations usually employed and to exhibit multimodality (Leistedt, Mortlock & Peiris 2016; see also Autenrieth et al. 2024, fig. 1). While this puts strain on likelihood-based analyses, SBI is completely transparent to the distributions used and so can easily handle realistic photometric redshifts.

On the other hand, the intensity of a supernova's light is affected by its distance $D$ from the observer: since the total flux is spread over a sphere with area $4\pi D^2$, the spectral flux *density* (SFD) is simply

$$F_o^s(t_o, \lambda_o) = \frac{\Phi_o^s(t_o, \lambda_o)}{4\pi D^2}. \quad (6)$$

For supernovæ located at cosmological distances, the appropriate distance to use (see e.g. Hogg 2000) is the *transverse comoving distance*[6] $D_M(z_c, \mathcal{C})$, which is related to the object's cosmological redshift $z_c^s$ through the cosmological model parametrized by $\mathcal{C}$.

Using SNæ Ia for cosmological inference therefore requires disentangling the effect of peculiar velocities[7] from the cosmological redshift, especially when using low-redshift ($z \lesssim 0.1$) SNæ Ia (Davis et al. 2011). While peculiar velocities can be included in the BHM and inferred simultaneously with other SN and host parameters (Rahman et al. 2022), cosmological analyses usually employ a simpler procedure of correcting the redshifts and propagating the associated uncertainty to the observed magnitudes. In Section 5, where we analyse real light curves, we will also use this simpler approach for comparison with previous work, while for the examination of simulated data in Section 4, we will assume no peculiar velocity, leading to an equality of all considered redshifts: $z_c^s = z^s = \hat{z}^s$. In future work, we will extend our SBI framework to perform joint inference of SNæ and their hosts, thus fully accounting for peculiar velocities.

### 2.4 Propagation effects: Milky Way extinction

Dust extinction in the Milky Way (MW) is similar to that in host galaxies, but the F99 law is evaluated at the observer-frame wavelengths:

$$F_o^s(t_o, \lambda_o) \rightarrow F_o^s(t_o, \lambda_o) \times [\text{F99}(\lambda_o; R_{V,\text{MW}})]^{A_{V,\text{MW}}^s}. \quad (7)$$

Following Schlafly & Finkbeiner (2011, hereafter SF11), we assume an isotropic MW dust law with $R_{V,\text{MW}} = 3.1$ and perfectly measured MW optical depths $A_{V,\text{MW}}^s$ at the sky locations of the SNæ, extracted from the SF11 maps. These assumptions can be easily relaxed and

$$F_o^s(t_o, \lambda_o) = \frac{1}{(1 + z_c)} \frac{\Phi_r^s[t_o/(1 + z_c), \lambda_o/(1 + z_c)]}{4\pi D_L^2},$$

where the prefactor $1/(1 + z_c)$ is cancelled when integrating to obtain bolometric fluxes. This formulation, however, makes dealing with peculiar velocities, which introduce redshift but not distance, cumbersome (Davis et al. 2011), and so we forward simulate the two effects separately.

---

[6]Alternatively, the combined effect of cosmological redshift and cosmological distance can be expressed through a *luminosity distance* $D_L = D_M(1 + z_c)$ as the familiar

[7]and other correlated sources of redshift like a local over/under-density (Wojtak, Davis & Wiis 2015)

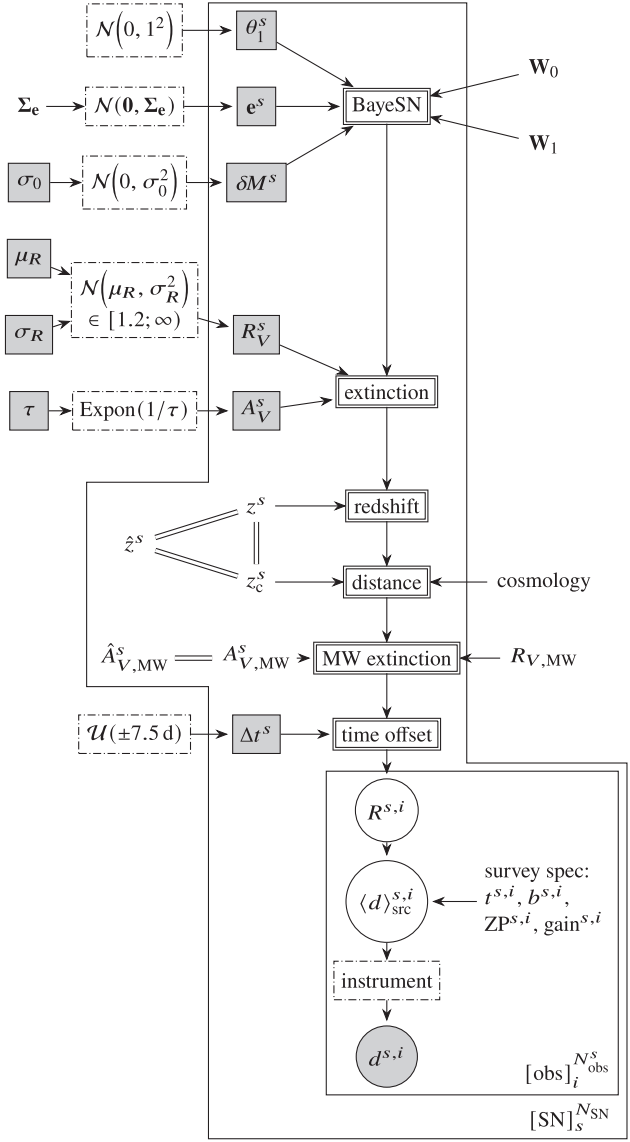

**Figure 1.** Graphical depiction of the model, in which parameters to be inferred are represented by shaded squares, the data by a shaded circle, and deterministic components and fixed variables by unshaded nodes. The distributions of SN-local variables (those inside the 'SN' plate) are in dashed boxes, while those for the global parameters are omitted for clarity. See Fig. 2 for an elaboration of the 'instrument' node and Table 1 for a tabular representation of the model.

the MW dust properties inferred or marginalized with SBI similarly to those of the hosts.

### 2.5 Instrument model

A telescope measures the *photon rate density* in a given band $b$ (in the observer frame):

$$R_b^s(t_o) = \int T_b(\lambda_o) \frac{F_o^s(t_o, \lambda_o)}{hc/\lambda_o} \, d\lambda_o, \tag{8}$$

where $T_b$ is the filter transmission (including the wavelength dependence of atmospheric absorption), and $hc/\lambda_o$ is an individual photon's (observer-frame) energy. In slicsim, the integral is performed numerically over a dense grid of wavelengths. Photometric light curves consist of a collection of noisy measurements of $R$ identified, in addition to the SN label $s$, by $i \in \{1, \ldots, N_{obs}^s\}$, for each of which a time $t^{s,i}$ and band $b^{s,i}$ are provided. The translation invariance of the time axis necessitates the introduction of a time-offset parameter, $\Delta t^s$, for each SN, so that $t^{s,i} = t_o^{s,i} + \Delta t^s$.

**Table 1.** SN Ia parameters, (hierarchical) priors and values used to generate mock data. For local parameters, the support and size of the sampled 'vector' are listed. See also Fig. 1 for a graphical representation of the model.

| parameter | | prior | mock value |
|---|---|---|---|
| BayeSN spectro-temporal templates ($k = 0, 1$) | $W_k$ | fixed | M20 |
| BayeSN 'stretch' parameter | $\theta_1^s$ | $\mathcal{N}(0, 1^2)$ | $\in \mathbb{R}^{\otimes N_{SN}}$ |
| BayeSN residual variations | $e^s$ | $\mathcal{N}(\mathbf{0}, \mathbf{\Sigma}_e)$ | $\in \mathbb{R}^{N_{grid} \otimes N_{SN}}$ |
| covariance of $e^s$ | $\mathbf{\Sigma}_e$ | fixed | M20 |
| abs. magnitude offset | $\delta M^s$ | $\mathcal{N}(0, \sigma_0^2)$ | $\in \mathbb{R}^{\otimes N_{SN}}$ |
| abs. magnitude scatter | $\sigma_0$ | HalfCauchy (0.1) | 0.088 |
| F99 dust law parameter | $R_V^s$ | $\mathcal{N}(\mu_R, \sigma_R^2)$ | $\in [1.2; \infty)^{\otimes N_{SN}}$ |
| 'mean' $R_V^s$ | $\mu_R$ | $\mathcal{U}(1.2, 5)$ | 3.0 |
| 'st. dev.' $R_V^s$ | $\sigma_R$ | HalfNormal($2^2$) | 0.5 |
| extinction optical depth | $A_V^s$ | Expon($1/\tau$) | $\in \mathbb{R}_+^{\otimes N_{SN}}$ |
| mean opt. depth | $\tau$ | HalfCauchy(1) | 0.329 |
| cosmological redshift | $z_c^s$ | fixed | $= z^s$ |
| total redshift | $z^s$ | fixed | $= \hat{z}^s$ |
| measured redshift | $\hat{z}^s$ | fixed | K17 |
| MW dust law parameter | $R_{V, MW}$ | fixed | 3.1 |
| MW optical depth | $A_{V, MW}^s$ | fixed | $= \hat{A}_{V, MW}^s$ |
| measured $A_{V, MW}^s$ | $\hat{A}_{V, MW}^s$ | fixed | SF11 |
| time offset | $\Delta t^s$ | $\mathcal{U}(\pm 7.5 \, d)$ | $\in [\pm 5 \, d]^{\otimes N_{SN}}$ |

The instrumental characteristics and amount of atmospheric absorption affecting the observation are summarized into a *zero-point magnitude*, which is the magnitude of a source which produces a read-out of 1 ADU on expectation. An ADU (analogue-to-digital unit), in turn, is related to photoelectrons, which are the actual Poisson-distributed quantity, through the instrumental *gain*. Finally, the magnitude system is defined through a *zero-point count (rate density)* $R_{ZP, b^{s,i}}$: this is the 'brightness', in the particular band, of an object of nought magnitude.

The expected signal (number of photoelectrons) from SN $s$ at time $t^{s,i}$ in band $b^{s,i}$ is, then

$$\langle d \rangle_{src}^{s,i} = \frac{R_{b^{s,i}}^s(t^{s,i})}{R_{ZP, b^{s,i}}} \frac{10^{0.4 \times ZP^{s,i}}}{gain^{s,i}}, \tag{9}$$

to which we must add background noise from: the electronics, the sky, and the host galaxy; cumulatively, $\langle d \rangle_{bg}$, before Poisson-sampling the final instrument readout:

$$d^{s,i} \sim \text{Pois}\left(\langle d \rangle_{src}^{s,i} + \langle d \rangle_{bg}^{s,i}\right). \tag{10}$$

While $R_{ZP, b^{s,i}}$ is measured in laboratory conditions, and $gain^{s,i}$ is a controllable setting, the zero-point magnitude includes contributions from the weather and other uncertain effects, so for each pointing $s, i$, the survey data release contains a *noisy measurement* of $ZP^{s,i}$, often represented through a mean zero point and a (usually small) zero-point uncertainty obtained from simultaneous observations of photometric standards.

**A common simplification** of the instrument description, which we will also adopt, is to consider a simple Gaussian model for the observed *calibrated flux* ('FLUXCAL') with mean $\langle d \rangle_{src}$ and standard deviation FLUXCALERR:

$$\text{FLUXCAL}^{s,i} \sim \left(\langle d \rangle_{src}^{s,i}, (\text{FLUXCALERR}^{s,i})^2\right) \tag{11}$$

In this case, linear rescaling does not modify inference, so ZP and gain can be set arbitrarily: to 27.5 and 1 as standardized by SNANA. The uncertainty, instead, is derived from data and combines in quadrature the (Gaussianized) uncertainties from the source, the electronics, the zero-point estimate, and from independent measurements of the sky and the host, which have been substracted[8] from the data to produce FLUXCAL.

If only FLUXCALs and FLUXCALERRs are released by a survey, instead of the more detailed information required to use equation (10), one could in principle simulate the noise sources, analyse them to produce 'mock' FLUXCALERRs, and include them as part of the *data* in an SBI framework. In this study, however, FLUXCALERRs are regarded as part of the instrument and are therefore kept fixed for simplicity, effectively assuming that Poisson noise from the source is a subdominant component, which justifies disregarding its dependence on the true source flux.

### 2.6 Hierarchical Bayesian modelling of SNæ Ia

The parameters that describe each individual supernova ($\theta_1^s$, $e^s$, $\delta M^s$), its environment ($R_V^s$, $A_V^s$), and other related quantities ($A_{V,\mathrm{MW}}^s$, $z^s$, $\Delta t^s$) are, in the context of a Bayesian hierarchical model (BHM), assigned priors which may themselves be parametrized by *global* hyperparameters to be jointly inferred with the SN-specific properties. This allows for so-called 'borrowing of strength': constraints from individual objects are partially pooled (with weights given by the relative uncertainty in each object), and the final posteriors – for both global *and* local parameters – benefit from the whole set of observations.

The hierarchical structure of our model is depicted in Fig. 1, and the variables and their (hyper)priors are listed in Table 1. We allow each SN Ia to be affected by dust with host-specific properties, adopting a hierarchical prior distribution of $R_V^s$ governed by hyperparameters as in Thorp et al. (2021), Thorp & Mandel (2022), Grayling et al. (2024). This hyperprior has the shape of a Gaussian, but its support is restricted to $[1.2; \infty)$ to maintain physicality (the lower limit is set by Rayleigh scattering; see Draine 2003). Thus, the hyperparameters $\mu_R$, $\sigma_R$ no longer represent the mean and standard deviation, respectively, of the distribution of $R_V^s$. Correspondingly, this truncation[9] modifies the likelihood of $\mu_R$, $\sigma_R$ by favouring broad $R_V^s$ distributions (i.e. a high $\sigma_R$), which would otherwise predict a number of objects with very low – or even negative – $R_V^s$ in contradiction with the data. Further discussion and explanation of the two effects can be found in Grayling et al. (2024).

For this proof-of-concept study, we adopt the following simplifications, which can easily be relaxed in future analyses. First, we assume perfect (spectroscopic) redshift estimates and no peculiar velocities, fixing $z^s = z_c^s = \hat{z}^s$ (except when analysing real data: see Section 5). We also fix the cosmological model[10] (and hence distances $D_\mathrm{M}^s$) to a flat $\Lambda$CDM with matter density $\Omega_{\mathrm{m}0} = 0.28$ and Hubble parameter $H_0 = 73.24 \, \mathrm{km \, s^{-1} \, Mpc^{-1}}$. We also assume perfectly measured MW optical depth, $A_{V,\mathrm{MW}}^s = \hat{A}_{V,\mathrm{MW}}^s$, and an isotropic fixed MW dust law with $R_{V,\mathrm{MW}} = 3.1$. Finally, we use a simplified Gaussian description of calibrated fluxes and a toy model for the uncertain time of maximum – a parameter that shifts the whole light curve rigidly in time – by allowing it to fall within a 15-day time window around an initial estimate (which we set as the time origin).

## 3 TRUNCATED MARGINAL NEURAL RATIO ESTIMATION

Consider a Bayesian model that includes parameters of interest $\boldsymbol{\Theta}$ and 'nuisance' parameters $\boldsymbol{\eta}$ in which we are generally not interested and would like to marginalize when analysing data $\boldsymbol{d}$. That is, we would like to obtain the marginal posterior

$$\mathrm{p}(\boldsymbol{\Theta}|\boldsymbol{d}) = \int \mathrm{p}(\boldsymbol{\Theta}, \boldsymbol{\eta}) \, d\boldsymbol{\eta} = \frac{\int \mathrm{p}(\boldsymbol{d}|\boldsymbol{\Theta}, \boldsymbol{\eta}) \mathrm{p}(\boldsymbol{\Theta}, \boldsymbol{\eta}) \, \mathrm{d}\boldsymbol{\eta}}{\mathrm{p}(\boldsymbol{d})}. \quad (12)$$

Marginal neural ratio estimation uses forward simulations to build a training set with two classes of $(\boldsymbol{\Theta}, \boldsymbol{d})$ pairs: either sampled from the joint distribution $\mathrm{p}(\boldsymbol{\Theta}, \boldsymbol{d})$, or from the product of marginals $\mathrm{p}(\boldsymbol{\Theta}) \, \mathrm{p}(\boldsymbol{d})$. The former are obtained by first running the full joint model $\mathrm{p}(\boldsymbol{\Theta}, \boldsymbol{\eta}, \boldsymbol{d})$ and simply disregarding $\boldsymbol{\eta}$, while for the latter, respectively, $\boldsymbol{d}$ or $\boldsymbol{\Theta}$ are further disregarded to have $\boldsymbol{\Theta} \sim \mathrm{p}(\boldsymbol{\Theta})$ and $\boldsymbol{d} \sim \mathrm{p}(\boldsymbol{d})$. A neural network $\hat{r}(\boldsymbol{\Theta}, \boldsymbol{d})$ is then trained to minimize the binary cross-entropy (BCE) loss:

$$\mathbb{E}_{\mathrm{p}(\boldsymbol{\Theta},\boldsymbol{d})}\left[-\ln\frac{\hat{r}(\boldsymbol{\Theta},\boldsymbol{d})}{1+\hat{r}(\boldsymbol{\Theta},\boldsymbol{d})}\right] + \mathbb{E}_{\mathrm{p}(\boldsymbol{\Theta})\mathrm{p}(\boldsymbol{d})}\left[-\ln\frac{1}{1+\hat{r}(\boldsymbol{\Theta},\boldsymbol{d})}\right], \quad (13)$$

which, as Hermans, Begy & Louppe (2020) show, leads to the neural network approximating the ratio

$$r(\boldsymbol{\Theta}, \boldsymbol{d}) \equiv \frac{\mathrm{p}(\boldsymbol{\Theta}, \boldsymbol{d})}{\mathrm{p}(\boldsymbol{\Theta})\mathrm{p}(\boldsymbol{d})} = \frac{\mathrm{p}(\boldsymbol{d}|\boldsymbol{\Theta})}{\mathrm{p}(\boldsymbol{d})} = \frac{\mathrm{p}(\boldsymbol{\Theta}|\boldsymbol{d})}{\mathrm{p}(\boldsymbol{\Theta})}. \quad (14)$$

An approximate posterior for the parameters of interest can then be obtained either by multiplying the prior density $\mathrm{p}(\boldsymbol{\Theta})$, if it is readily available, by $\hat{r}(\boldsymbol{\Theta}, \boldsymbol{d}_0)$, evaluated at the observed (as opposed to simulated) data $\boldsymbol{d}_0$, or more generally, by weighting prior samples, e.g. those used for training/validation, by $\hat{r}(\boldsymbol{\Theta}, \boldsymbol{d}_0)$.

NRE is an amortized technique, which means that the NN tries to learn $r(\boldsymbol{\Theta}, \boldsymbol{d})$ even for data that does not resemble the observed one. While this allows its perfromance to be tested and verified, it is usually advisable to re-train using simulations focused on the particular (target) observation $\boldsymbol{d}_0$. We use the truncation scheme of Miller et al. (2021), whereby global-parameter priors are restricted to regions with significant posterior mass (approximately determined via an initial NRE run), but their shapes are not modified. This preserves the amortization of the ratio estimator within the confines of the truncated priors. We found that truncation was only necessary for the most complicated inference tasks (global dust population parameters), while other posteriors were optimally recovered from their initial priors.

### 3.1 Inference network

An SBI analysis is only as powerful as the inference network employed in it is expressive. A collection of light curves is a peculiar data set since the data $\boldsymbol{d}^s \equiv [d^{s,i}]_{i=1}^{N_{\mathrm{obs}}^s}$ related to each object have different sizes because of the irregular cadence of observations. While NN architectures that accept varying-length inputs exist[11] and are presently ubiquitous in areas like natural language processing, in the present work, we use a simpler architecture that we found is fast to train (both in terms of number of steps and time per step) and with present-day data sets achieves the best performance while requiring reasonable resources.

---

[8] Of course, this might produce negative (calibrated) 'fluxes' (which are not even fluxes but photon counts), but one learns to live with it.

[9] Curiously, although not part of the truncation scheme for neural ratio estimation that we use to improve the training and performance of our inference network, the likelihood modification discussed here also applies to – and is a pitfall for – hierarchical TMNRE.

[10] Since we simulate a low-redshift survey (see Section 4.1), we have little hope of inferring cosmology beyond the combination of the Hubble parameter and the SN Ia average absolute magnitude. We choose to fix cosmology for now in order to allow comparison with the existing likelihood-based BayeSN inference code (see Section 4.3), although relaxing this is the principal goal of our future work.

[11] For SBI applications, see Rodrigues et al. (2021), Campeau-Poirier et al. (2023), Heinrich et al. (2024).

As a key first step, we use a collection of $N_{SN}$ *distinct* learnable embeddings of the unequal-length $d^s$ into a space of fixed dimension:

$$\boldsymbol{f}^s \equiv \texttt{SNEmbed}_s(d^s). \quad (15)$$

We found that even a single linear layer (per SN) works well in our particular set-up. The embeddings are stacked along a batch dimension and processed in parallel by a single *shared* component to derive final featurized representations of each supernova:

$$\boldsymbol{d}^s \equiv \texttt{SNHead}(\boldsymbol{f}^s). \quad (16)$$

A data set summarizer then combines information from all objects:

$$\boldsymbol{\mathcal{S}} \equiv \texttt{Summariser}\left([\boldsymbol{d}^s]_{s=1}^{N_{SN}}\right). \quad (17)$$

We use a simple summarizer that concatenates the (ordered) tuple of inputs $[\boldsymbol{d}^s]_{s=1}^{N_{SN}}$, as we previously did in SICRET, where we showed that, while memory- and compute-intensive, this approach does scale to $\sim 10^5$ SNæ with current hardware. We prevent overfitting in this layer via stochastic dropout (Hinton et al. 2012).

Finally, a number of ratio-estimator networks estimate ratios. For a group of global parameters $\boldsymbol{\Theta}$:

global: $\quad \ln \hat{r}(\boldsymbol{\Theta}, d) = \texttt{RatioEstimator}_\Theta(\boldsymbol{\Theta}, \boldsymbol{\mathcal{S}}), \quad (18)$

and similarly for the local parameters $\theta^s$ of object $s$:

local: $\quad \ln \hat{r}(\theta^s, d) = \texttt{RatioEstimator}_\theta(\theta^s, \boldsymbol{\mathcal{S}}, d^s, \mathbf{a}^s), \quad (19)$

where $\mathbf{a}^s \equiv [\hat{z}^s, \hat{A}^s_{V,\text{MW}}]$ are auxiliary (constant) inputs that completely identify the object whose parameters are being inferred.[12] As we previously noted in SICRET, the presence of $\boldsymbol{\mathcal{S}}$ accounts for *a posteriori* correlations between the parameters: i.e. we infer the posterior $p(\theta^s|\{d^s\})$ instead of $p(\theta^s|d^s)$. In a hierarchical model in which $\{\theta^s\}$ are *a priori* conditionally independent given global parameters $\boldsymbol{\Theta}$, i.e. $p(\{\theta^s\}|\boldsymbol{\Theta}) = \prod_s p(\theta^s|\boldsymbol{\Theta})$, this corresponds to the marginalization $\int p(\theta^s, \boldsymbol{\Theta}|d) d\boldsymbol{\Theta}$ instead of simply $p(\theta^s|d, \boldsymbol{\Theta})$ as was done in previous hierarchical SBI analyses of permutation-invariant data (Rodrigues et al. 2021; Heinrich et al. 2024).

To enhance their expressivity, ratio estimators first 'featurize' the raw parameters by passing them through a $\texttt{ParamHead}_\Theta$, whose output $\boldsymbol{\Theta}$ is concatenated to the data set summary. For latent-variable estimators, $\theta^s \equiv \texttt{ParamHead}_\theta(\theta^s)$ is concatenated with a processed $\boldsymbol{\mathcal{S}}_\theta \equiv \texttt{SummaryHead}_\theta(\boldsymbol{\mathcal{S}})$, which extracts the relevant summaries from $\boldsymbol{\mathcal{S}}$, with the pre-processed data $d^s$, and with the auxiliary inputs $\mathbf{a}^s$. Lastly, to enhance constraining power when the posterior is significantly more concentrated than the prior, we use the leaky parity-odd power (POP) activation layer (Jeffrey & Wandelt 2024) on the output of global-parameter ratio estimators. All network components (SN embedders, summarizer, and all ratio estimators) are trained simultaneously using the loss from equation (13) summed over all marginal parameter groups (i.e. $\boldsymbol{\Theta}$ representing, in turn, each of the global parameters and each of the local parameters, summing over the SNæ).

We implement all the components using multilayer perceptrons

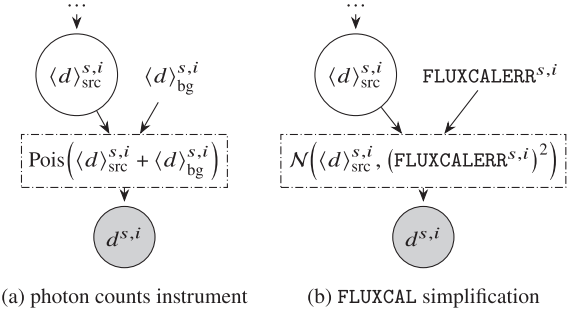

**Figure 2.** Two kinds of instrument, as discussed in Section 2.5. In this work, we only use the 'FLUXCAL' simplification from Fig. 2(b) but include Fig. 2(a) as a more principled approach we will adopt in the future.

(MLPs) with batch normalization and rectified linear unit (ReLU) non-linearities. Details about layer sizes are given in Table 2, and the network is depicted in Fig. 3.

## 4 DEMONSTRATION ON SIMULATED DATA

### 4.1 Mock data and survey specification

We generate and analyse mock data designed to mimic the third data release of CSP, as presented in Krisciunas et al. (2017) and included in SNANA (Kessler et al. 2009). We extract the list of observation times ($t^{s,i}$) and bands ($b^{s,i}$) for each of the SNæ Ia in the data release, their spectroscopic redshift ($\hat{z}^s$) and the Milky Way colour excess $E(B-V) \equiv A_B - A_V$, which we convert into $\hat{A}^s_{V,\text{MW}} = 3.1 \times E(B-V)$ (since we assume isotropic $R_V = 3.1$ for the Milky Way). This constitutes the 'survey specification' part of the input to the graphical model from Fig. 1.

Since the M20 model, which we use both to generate and analyse the mock data, was not trained on *u*-band observations and outside the rest-frame time range $[-10\,\text{d}; 40\,\text{d}]$, we exclude the corresponding entries from our CSP-like set-up. As described in Section 2.1, we keep the spectro-temporal templates $\boldsymbol{W}_k$ and the covariance $\boldsymbol{\Sigma}_e$ of intrinsic residual variations fixed in this work (they are inputs to the graphical model of Fig. 1). For the remaining parameters, we set ground-truth values as listed in Table 1, informed by the posterior means reported in M20. Individual light curves are simulated as described in Section 2 after sampling SN-local parameters from their respective priors (also listed in Table 1 and depicted in Fig. 1). We also add a random time shift of up to 5 d (rest-frame) around the time of the brightest observation (SEARCH_PEAKMJD) while, in order to avoid boundary effects in the analysis, we widen the prior to $\mathcal{U}(\pm 7.5\,\text{d})$.

In this work, we only simulate and analyse FLUXCAL data since detailed descriptions of the CSP observations (zero-points, gains, and background fluxes) are not available in SNANA. To facilitate the likelihood-based comparison, we increase very small reported FLUXCALERRs to be at least 0.01 mag, as has also been done for the real data.

The mock data set contains $N_{SN} = 134$ simulated SNæ Ia with a total of $\sum_{s=1}^{N_{SN}} N^s_{\text{obs}} = 13\,202$ flux measurements.[13] Fig. 4 depicts it four times, coloured according to the values of the different SN-local parameters. The impact of $A_V$ and $\theta_1$ are clearly evident as gradients (shifts) of the light curves, while those of $\delta M$ and $R_V$ less so, which has an impact on the inference of the parameters in question. Notice also the reduced spread of SN Ia light curves in the infrared bands, where measurements allow disentangling pure-magnitude from colour-and-magnitude variations (respectively described by $\delta M$ and $R_V$).

---
[12]We streamline the discussion and notation from SICRET, section 3, and directly present the case where the same NN is used as the $\theta$-ratio estimator for each object, which necessitates the use of $\mathbf{a}^s$.

[13]For the mock data that we generate ourselves, we consider all SNæ from the CSP data release instead of the restricted 'clean' sample we analyse in Section 5.

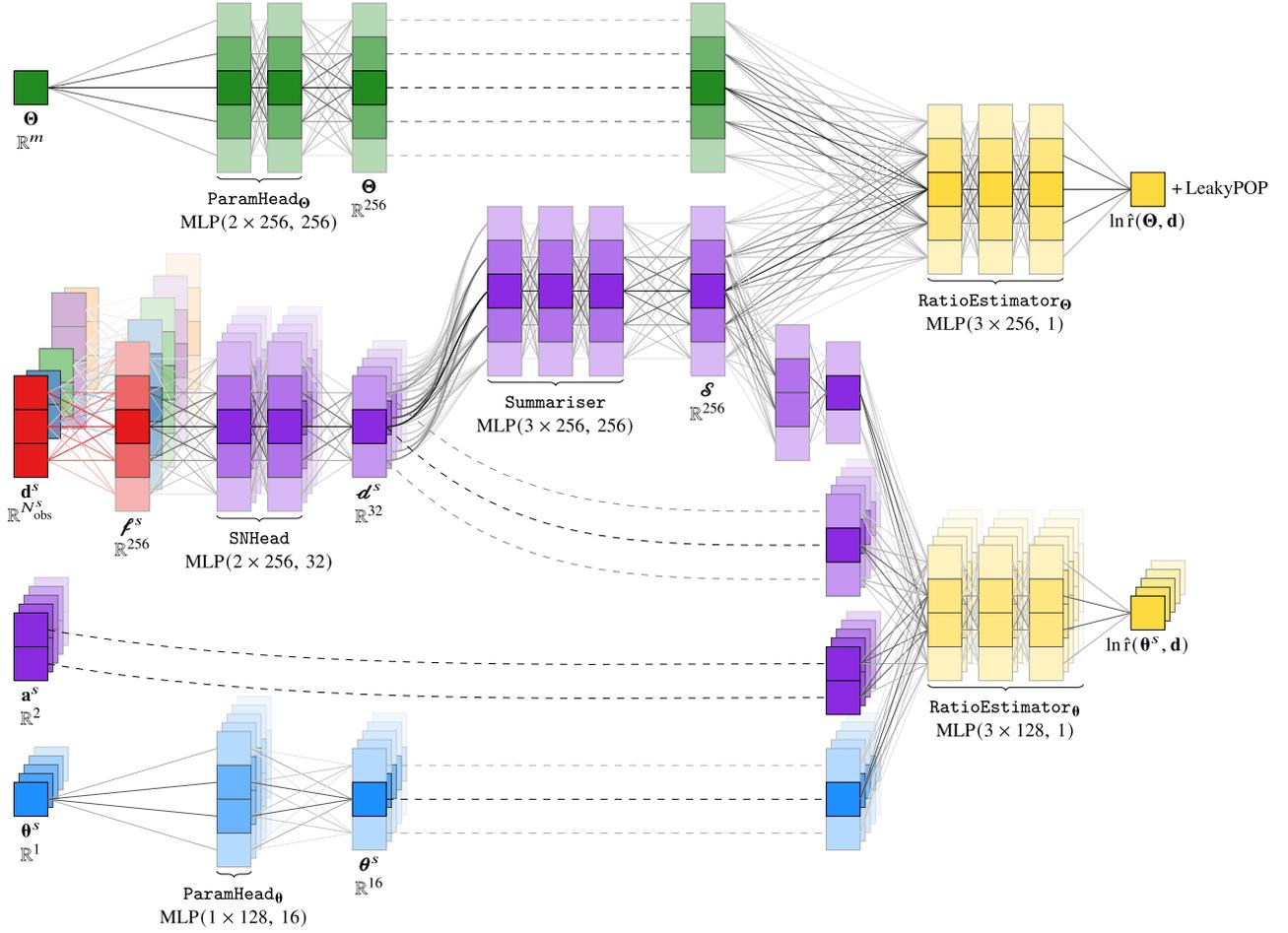

**Figure 3.** Architecture of the neural network. Solid lines represent linear connections (followed inside the MLPs by batch normalization and ReLU non-linearity). Dashed lines, on the other hand, connect layers that are duplicated for presentation (identity operation). When multiple parameter groups are being inferred (at the initial stage before truncation), there are multiple parameter heads, whereas here only one is shown for clarity. Notice the similarity with SICRET, fig. 2, the main addition being the SN-embedding layers: $N_{\rm SN}$ distinct components (thus coloured diversely) with different input sizes but the same output dimension, which allow the $\{\boldsymbol{f}^s\}$ to be stacked into a single tensor along a batch dimension and processed in parallel before being flattened out for input into the summarizer.

**Table 2.** Details about the components of the inference network: their input and output dimensions and particular implementation in this work. For all components, we use MLPs, indicating the number and size of the hidden layers and the output size as MLP($n_{\rm hidden} \times d_{\rm hidden}$, $d_{\rm out}$). Each hidden layer consists of a fully connected layer, batch normalization, and a ReLU non-linearity. Inputs are also whitened (shifted and scaled by the mean and standard deviation of the training set). The size of global-parameter groups is denoted with $m = 1$, or 2 for the group $[\mu_R, \sigma_R]$. The inference network is also depicted in Fig. 3.

| component | inputs | $\in$ | space | $\to$ | output | $\in$ | space | implementation |
|---|---|---|---|---|---|---|---|---|
| $[\texttt{SNEmbed}_s]_{s=1}^{N_{\rm SN}}$ | $\boldsymbol{d}^s$ | $\in$ | $\mathbb{R}^{N^s_{\rm obs}}$ | $\to$ | $\boldsymbol{f}^s$ | $\in$ | $\mathbb{R}^{256}$ | Linear($N^s_{\rm obs} \to 256$) |
| $\texttt{SNHead}$ | $\boldsymbol{f}^s$ | $\in$ | $\mathbb{R}^{256}$ | $\to$ | $\boldsymbol{d}^s$ | $\in$ | $\mathbb{R}^{32}$ | MLP($2 \times 256$, 32) |
| $\texttt{Summariser}$ | $[\boldsymbol{d}^s]$ | $\in$ | $\mathbb{R}^{32 \times N_{\rm SN}}$ | $\to$ | $\mathcal{S}$ | $\in$ | $\mathbb{R}^{256}$ | MLP($3 \times 256$, 256) |
| $\texttt{ParamHead}_\Theta$ | $\Theta$ | $\in$ | $\mathbb{R}^m$ | $\to$ | $\Theta$ | $\in$ | $\mathbb{R}^{256}$ | MLP($2 \times 256$, 256) |
| $\texttt{RatioEstimator}_\Theta$ | $\Theta, \mathcal{S}$ | $\in$ | $\mathbb{R}^{256+256}$ | $\to$ | $\ln \hat{r}(\Theta, d)$ | $\in$ | $\mathbb{R}^1$ | MLP($3 \times 256$, 1) + LeakyPOP |
| $\texttt{ParamHead}_\theta$ | $\theta^s$ | $\in$ | $\mathbb{R}^1$ | $\to$ | $\theta^s$ | $\in$ | $\mathbb{R}^{16}$ | MLP($1 \times 128$, 16) |
| $\texttt{SummaryHead}_\theta$ | $\mathcal{S}$ | $\in$ | $\mathbb{R}^{256}$ | $\to$ | $\mathcal{S}_\theta$ | $\in$ | $\mathbb{R}^{16}$ | MLP($1 \times 128$, 16) |
| $\texttt{RatioEstimator}_\theta$ | $\theta^s, \mathcal{S}_\theta, \boldsymbol{d}^s, \boldsymbol{a}^s$ | $\in$ | $\mathbb{R}^{16+16+32+2}$ | $\to$ | $\ln \hat{r}(\theta^s, d)$ | $\in$ | $\mathbb{R}^1$ | MLP($3 \times 128$, 1) |

### 4.2 Training

We implement and train the neural network described in Section 3.1 in PyTorch (Paszke et al. 2019) and PyTorch Lightning (Falcon & The PyTorch Lightning team 2023). We target multiple parameter groups – global $\Theta_i \in \{\tau, \sigma_0, [\mu_R, \sigma_R]\}$ and local $\theta^s_i \in$ $\{\Delta t^s, \theta_1^s, A_V^s, \delta M^s, R_V^s\}$ – simultaneously by training separate ratio estimators for each at the same time as a single data pre-processor $\mathbf{d} \to \left[\{\boldsymbol{d}^s\}_{s=1}^{N_{\rm SN}}, \boldsymbol{\mathcal{S}}\right]$.

We generate a training set of 256 000 mock survey realizations and use 6400 additional examples for validation, plotting posteriors, and calibration. To prevent overfitting, while training, we resample the instrumental noise (equation (11)), which effectively augments the training set but avoids the expensive part of the simulator, and stochastically 'drop out' 50 per cent of the summarizer input (Hinton et al. 2012). We optimize using Adam (Kingma & Ba 2017) with the default PyTorch momentum settings, a decaying learning rate schedule ($\gamma = 1/1.5$ every 10 epochs) over 100 epochs, and with mini-batch size of 128 examples. The results we present below use the checkpoint that performed best on the validation set. Training on one NVIDIA A100 GPU took $\approx 5\,\mathrm{h}$ to converge, in addition to $\approx 30$ min needed to generate the training set. Evaluating a single set of marginal posteriors then takes on the order of milliseconds.

### 4.3 Validation with HMC

To validate our NRE results, we run a likelihood-based analysis (using the hierarchical likelihood that corresponds to the forward model in Section 2) with Hamiltonian Monte Carlo (HMC) and consider the resulting posterior the ground truth. We use the code outlined in Grayling et al. (2024), which is based on the implementation of the No-U-Turn Sampler (NUTS; Hoffman & Gelman 2014) in NumPyro (Bingham et al. 2019; Phan, Pradhan & Jankowiak 2019). We run four chains and draw 500 samples each after 500 burn-in steps, which takes $\approx 30$ min when run in parallel on four NVIDIA A100 GPUs. We verify convergence using the Gelmann–Rubin $R$ number, the effective sample size, and other standard diagnostics as described in Grayling et al. (2024). We remind the reader that, as any likelihood-based method, this HMC analysis requires sampling the joint posterior of all model parameters, including in this case four global parameters and $N_{\rm SN} \times (5 + 42) = 6298$ object-specific ones, most of which describe the residual light-curve variations through $\boldsymbol{e}^s$. In contrast, our SBI methodology implicitly marginalizes $\boldsymbol{e}^s$ and estimates three global (since we group $[\mu_R, \sigma_R]$) and $N_{\rm SN} \times 5$ SN-specific marginal posteriors.

### 4.4 Comparison of marginal posteriors

We plot marginal NRE posteriors, evaluated by weighting prior samples from the validation set by $\hat{r}(\boldsymbol{\Theta}, \boldsymbol{d})$ in Fig. 5, compared with the ground-truth marginalized HMC posteriors and the true parameter values used to generate the mock data.

We observe excellent agreement between NRE and HMC posteriors for the global parameters: $\tau$, $\sigma_0$, $[\mu_R, \sigma_R]$, with similar uncertainties from the two methods and relative shifts of at most about $1\sigma$. Since the ratio estimator for $[\mu_R, \sigma_R]$ is the most challenging, we re-trained it after truncating the global-parameter priors, as described by Miller et al. (2021) and illustrated in the figures.

Similarly, SN-local parameters, with the exception of $R_V^s$, are very well recovered and in agreement with HMC. A detailed comparison of the first two moments of the one-dimensional marginal posteriors is shown in the top row of Fig. 5. In general, NRE exhibits slightly larger uncertainties for most parameters, as was previously observed in SICRET. It is important to note that $R_V^s$ inference is almost entirely population-driven. Since constraints from individual objects are weak, the hierarchical structure induces 'shrinkage': a statistical effect whereby all *marginal* $R_V^s$ posteriors concentrate toward the population mean. This is not an artefact of the inference procedure used but rather a feature of the hierarchical model itself, and is observed for both NRE and HMC. Thus, small changes in the $\mu_R$–$\sigma_R$ posterior shift all the $R_V^s$ marginals coherently, leading to similar offsets from the HMC results for individual objects. We note that, while the $N_{\rm SN}$ + two-dimensional joint $\mu_R - \sigma_R - \{R_V^s\}_{s=1}^{N_{\rm SN}}$ posterior can be studied using HMC, with NRE, we only derive marginal posteriors.

## 5 APPLICATION TO REAL DATA

We apply the methodology described and validated above to the real light curves from CSP: specifically, the subset of 86 non-peculiar ones identified and analysed by Thorp & Mandel (2022, hereafter, TM22). Since this is a low-redshift sample, we use redshifts corrected for peculiar velocity (using the flow model of Carrick et al. 2015, as described in TM22) and thus have separate (fixed) $z^s$ and $z_c^s$, the former acting to redshift the light curves, while the latter is only used to calculate distances under a fixed cosmological model (flat $\Lambda$CDM with $\Omega_{\rm m0} = 0.28$ and $H_0 = 73.24\,\mathrm{km\,s^{-1}\,Mpc^{-1}}$). As standard in SN Ia analyses, we account for a $\pm 150\,\mathrm{km\,s^{-1}}$ uncertainty in the peculiar velocity correction by propagating it linearly to magnitudes[14]:

$$\sigma_\mu^s = \frac{5}{\ln 10} \frac{1}{\hat{z}^s} \sqrt{\left(\frac{150\,\mathrm{km\,s^{-1}}}{c}\right)^2 + \left(\sigma_z^s\right)^2}, \quad (20)$$

where $c$ is the speed of light, and $\sigma_z^s$ is the measurement uncertainty (which is small for spectroscopic redshift estimates). This is then included in the BHM as an additional (in quadrature) spread of absolute magnitudes: $\delta M^s \sim \mathcal{N}(0, \sigma_0^2 + (\sigma_\mu^s)^2)$. To further match TM22's set-up, we also fix the standard deviation of residual scatter $\sigma_0 = 0.088$ and the time offsets $\Delta t^s = 0$ instead of inferring them.

We present the posterior for $\mu_R$ and $\sigma_R$ in Fig. 6 in comparison with the one from TM22. A full comparison is shown in Fig. 7. The NRE- and HMC-derived posteriors are in good agreement with about $1\sigma$ offset and similar sizes, as was the case when analysing the simulated data set. Moreover, for completeness in Appendix A, we validate the global-parameter inference on simulated data with parameters randomly drawn from the priors, observing good Bayesian coverage properties, and calibrate the approximate posteriors to construct confidence regions with exact frequentist coverage. Since the NRE is already nearly optimal, this procedure does not lead to results significantly different from those presented in Figs 7 and 6 but serves as reassurance of their correctness.

## 6 DISCUSSION, OUTLOOK, AND CONCLUSION

We have presented detailed marginal neural simulation-based inference in the context of a hierarchical model of SN Ia light curves that incorporates realistic intrinsic (to each SN) and extrinsic (due to dust properties of the host galaxy) variability. By training a neural network to approximate the likelihood-to-evidence ratio with a training set of simulated light curves based on the Carnegie Supernova Project (CSP), we have derived marginal posteriors for the parameters of the populations of SNæ Ia and their hosts: the mean and standard deviation of dust-law parameters $R_V^s$, the average optical depth $\tau$, and the residual scatter of SN Ia absolute magnitudes $\sigma_0$, and simultaneously inferred marginally the parameters of all $\approx 100$ SNæ Ia. After validating the approach on simulated data, we have analysed the light curves of 86 real SNæ Ia observed by the CSP (Krisciunas et al. 2017) and selected by TM22. In both cases, we observe excellent agreement between our SBI results and a baseline likelihood-based analysis as in Thorp & Mandel (2022), Grayling et al. (2024). Concretely, posteriors for $\tau$ and $\sigma_0$ are in perfect agreement from the two methods, as well as the marginals for most local parameters ($\Delta t^s$, $\theta_1^s$, $A_V^s$, $\delta M^s$). For the latter, SBI exhibits a slightly bigger uncertainty (by $\approx 10$ per cent). Results for the dust-law parameters $R_V^s$ and their population are also in good agreement, with only a small offset of about $1\sigma$ between NRE and HMC observed. As we illustrate in Figs 4 and 8, $R_V^s$ have a minuscule impact on the data in comparison with the remaining variability, which makes them the hardest to infer and leads to hierarchy-dominated results: i.e. inference of one $R_V^s$ depends on observations of all SNæ, regardless of the analysis methodology (likelihood- or simulation-based). In light of this extremely challenging learning task, the neural network exhibits excellent performance, having learned to extract and route the relevant information without access to the full high-dimensional likelihood but solely from training examples.

---

[14] As discussed in Grayling et al. (2024), dust inference is only mildly affected by the peculiar velocity model, which mainly trades off against the residual scatter $\sigma_0$. However, in SICRET, we showed that standard error propagation through the non-linear distance modulus can significantly bias cosmological inference from large collections of SNæ Ia. With SBI, it would be trivial to avoid this approximation and instead correctly account for (and marginalize) any redshift uncertainties, including those from photometric fits, on top of a spread of peculiar velocities consistent with large-scale galaxy-flow models (see, e.g. Rahman et al. 2022). Here, we use the linear approximation solely for consistency with likelihood-based codes.

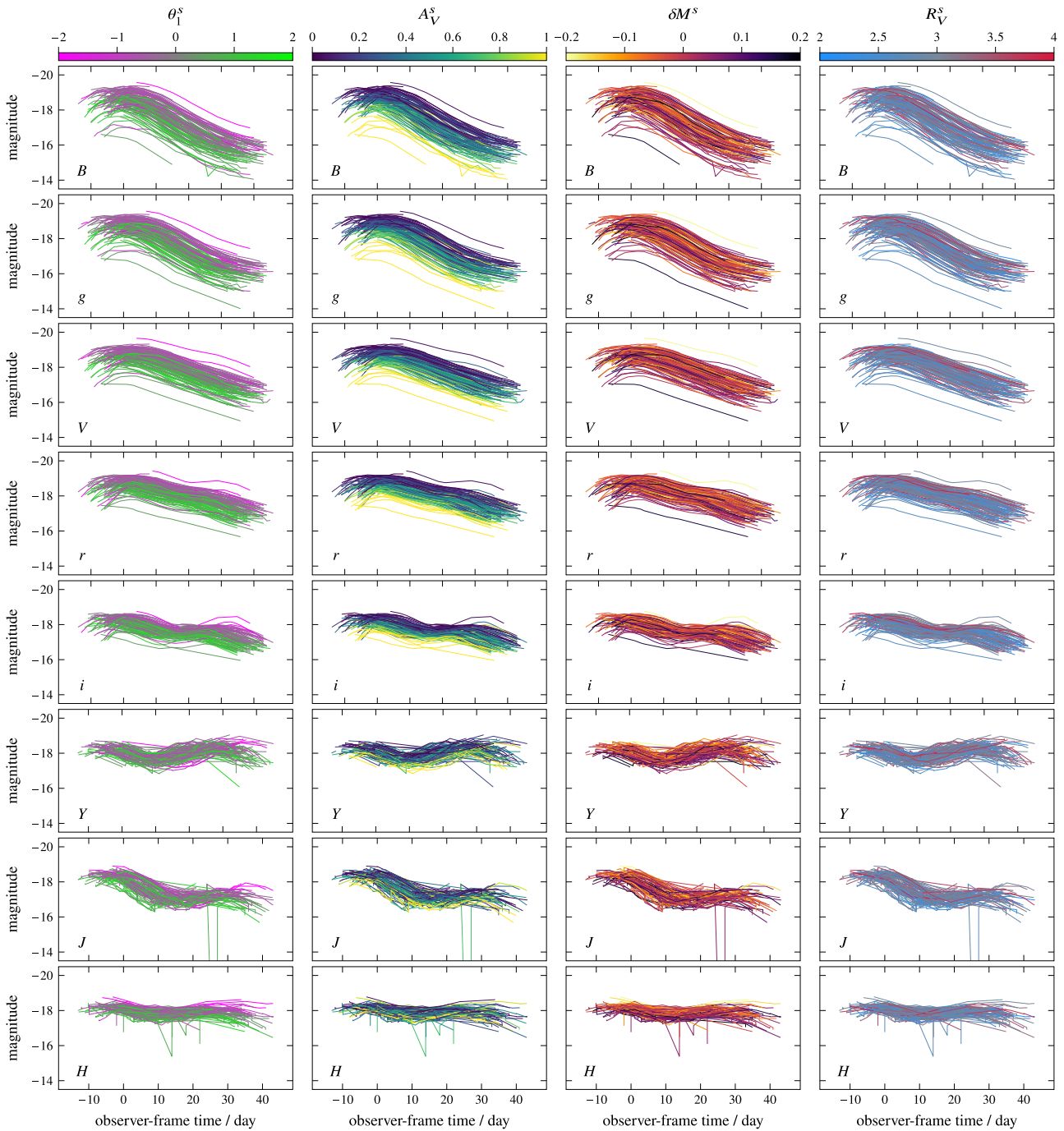

**Figure 4.** The mock light curves that we generate and analyse, corrected for cosmological distance (but not for redshift). While we work entirely in linear (flux) scale, for presentation purposes, this figure is in magnitudes. Each column shows the same light curves but coloured according to a different SN-local variable, as indicated on the top. Each row is a different CSP band: from bluest (top) to (infra)reddest (bottom). Different SNæ might have observations in different sets of the bands. All plots have the same scale and limits: notice that the diversity in redder bands is smaller, owing partly to the weaker effect of dust extinction.

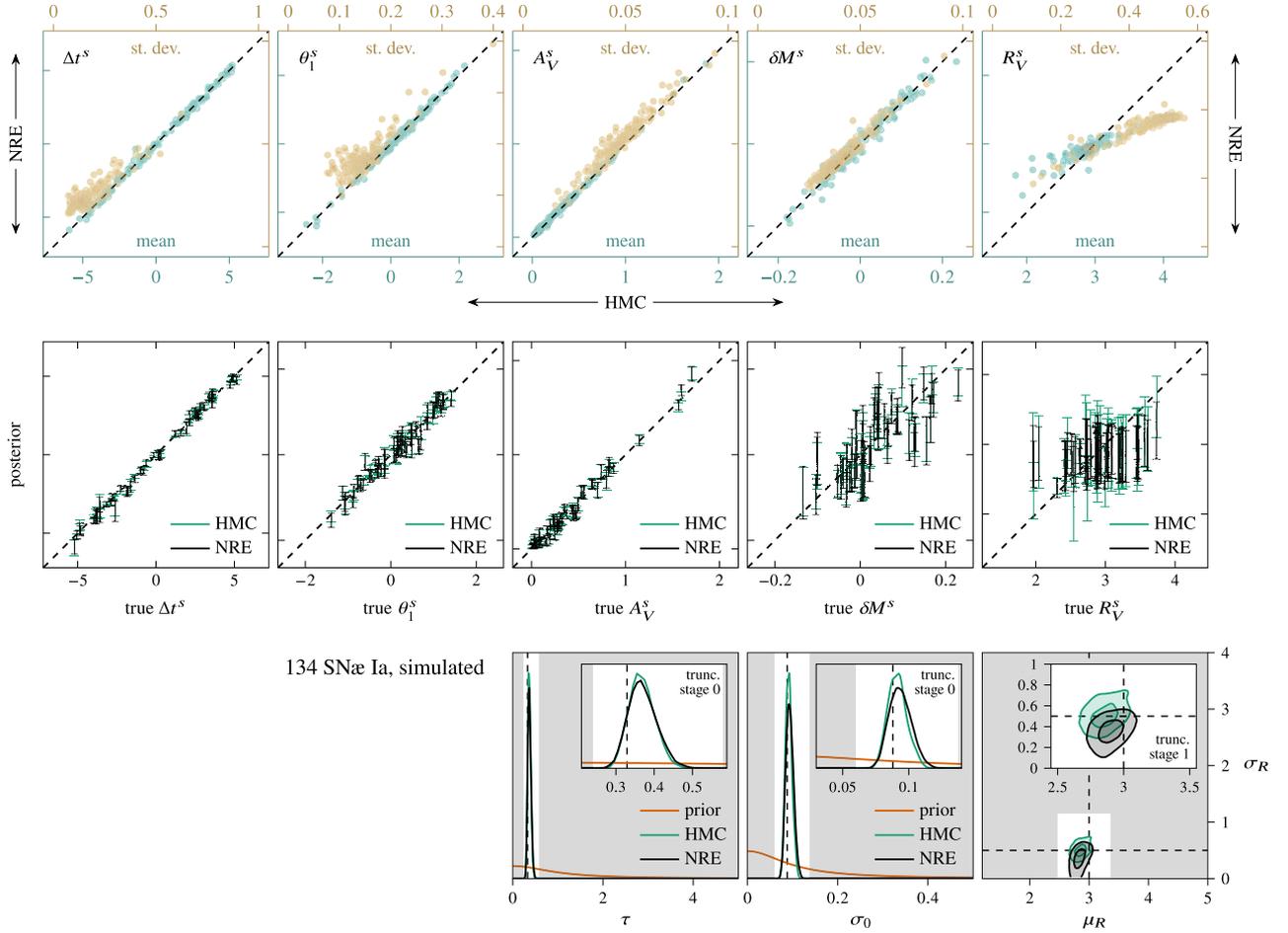

**Figure 5.** Inference results from the mock data set. *Top*: moments of the marginal posteriors of the local parameters of the $N_{\rm SN}$ supernovæ, as indicated in the top-left corner of each plot. Means (standard deviations) are shown in teal (ochre), with scale indicated below (above) the plot. The abscissa (ordinate) coordinate comes from the HMC (NRE) posterior, so that the diagonal indicates matching moments from the two methods. *Middle*: the same per-object marginal posteriors (mean $\pm 1$ standard deviation) plotted against the true values in the simulation. Only every third error bar is plotted for clarity. *Bottom*: posterior densities (in the two-dimensional plot, 1- and $2\sigma$ credible regions) for the global parameters, as inferred by HMC and NRE, compared with the prior density and the true value used to simulate the mock data. Shaded regions indicate the truncation used for re-training the $\mu_R$–$\sigma_R$ NRE posterior depicted in the inset (the estimators for $\tau$ and $\sigma_0$ were not re-trained).

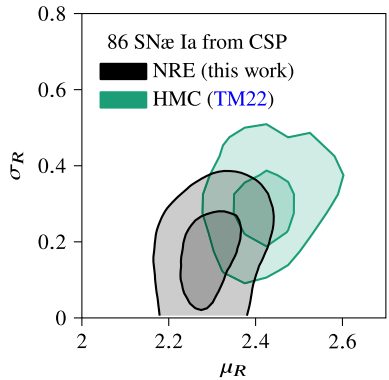

**Figure 6.** Comparison of marginal dust-population posteriors (1- and $2\sigma$ credible regions) from the subset of 86 SNæ Ia analysed in TM22 (assuming no split according to host mass: upper left in fig. 8 of TM22). The small offset is comparable to the one observed with mock data (Fig. 5) and is due to the minuscule effect that varying $R_V^s$ has on the data (see Figs 4 and 8) and the hierarchical nature of $\mu_R$–$\sigma_R$ inference, which makes the problem particularly challenging.

The precision and accuracy we achieve are largely due to the network architecture we adopt, designed to address the issue that each supernova in a survey has a different number of observations at different times and in different bands (thus, a survey is a *tuple*: an ordered collection of different-sized objects). Our NN consists of: a single bespoke linear layer embedding each SN individually into a common-dimensional space; a *shared* fully connected SN post-processing subnetwork applied in parallel to the embeddings of all SNæ; and a fully connected summarizer combining the results. It is as expressive and fast to evaluate and train as conventional fully connected networks (taking a few hours to converge with training data generated in ∼30 mins, in the same ballpark as highly optimized likelihood codes) but manages to fully extract the relevant information before overfitting.

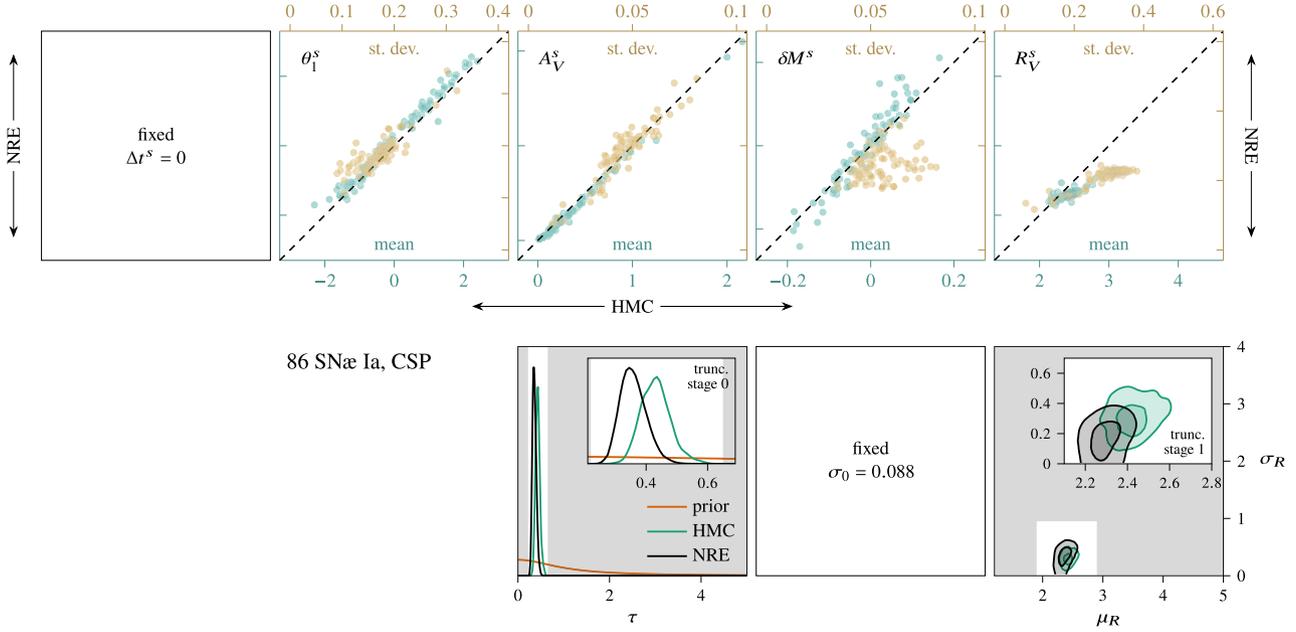

**Figure 7.** Inference results from the real data set. *Top*: comparison of marginal local-parameter posterior moments derived with NRE and HMC. *Bottom*: posteriors for the global parameters. See Fig. 5 for more details.

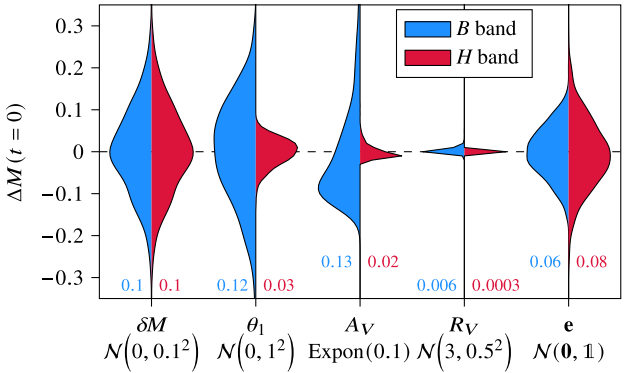

**Figure 8.** Variations in rest-frame *B* and *H* absolute magnitudes at phase 0 (around maximum), as simulated with the BayeSN trained by M20, induced by varying each of the free local parameters according to its fiducial hierarchical prior, with respect to a reference value with $\delta M = 0$, $\theta_1 = 0$, $A_V = 0.1$, $R_V = 3$, $\epsilon = 0$. Numbers along the bottom specify the standard deviation for the two bands.

In the present work, we have made a number of simplifying assumptions that do not affect significantly the results in the low-redshift, fairly small-size, high-signal-to-noise case we consider. For example, we employ the simplified instrumental model that summarizes observational uncertainties in a FLUXCALERR; to properly investigate the impact of measurement-related systematics, one must introduce calibration parameters (e.g. $ZP^{s,i}$) for each data point, which would significantly impact runtimes of likelihood-based methods.[15] The same applies to the hierarchical parameters we have kept fixed in this study: namely, the MW dust law (and its variation along different lines of sight) and extinction amount and the total and cosmological redshifts of the supernovæ. In contrast, with SBI all extra parameters can be marginalized implicitly by stochastically sampling them in the simulator, or marginally inferred in the same way (and at the same time) as the parameters we have already considered, provided suitable models: e.g. Zhang, Yuan & Chen (2023) for MW dust and Rahman et al. (2022) for redshifts induced by peculiar velocities consistent with models of the galactic bulk flow.

Along the same line of thought, one can also consider training the SN Ia template used to simulate light curves – represented in BayeSN through $W_k$ and $\Sigma_e$ – together with the properties of individual SNæ Ia (notably, distances used for cosmology). Even though this is the main reason behind using a hierarchical model, for BayeSN this is yet to be attempted because of subtleties related to selection effects. Likelihood-based analyses are thus typically split in two stages, with the second one (cosmological inference) using a fixed mean/median SN Ia spectral energy distribution (SED) model. To SBI, on the other hand, $W_k$ and $\Sigma_e$ are just another set of global parameters implicitly marginalized in the training data. Sampling them from the very general priors used by M22 (and successors) will introduce tremendous and difficult-to-handle variability in the training data, which, however, can be remedied through high-dimensional truncated SBI (see, e.g. Anau Montel, Alvey & Weniger 2023; List, Montel & Weniger 2023). Furthermore, SBI opens up the possibility of using *implicit* (data-driven) priors and/or SN templates implemented as neural generative models trained (or fine-tuned)

---

[15]Current state-of-the-art analyses (e.g. Brout et al. 2022a, b; Vincenzi et al. 2024) employ a simplistic yet laborious procedure of linear propagation of systematic 'uncertainties' (in fact, offsets) to the final results (e.g. estimates of cosmological parameters). This can be streamlined and made more principled with the SBI approach introduced here.

simultaneously with the inference, as recently done by Alsing et al. (2024) for galaxy photometry.

Beyond individual objects, numerous qualitative enhancements of the population modelling of SNæ Ia and their hosts have been explored in the literature. These follow two main threads: considering correlations of SN properties with those of their hosts, and allowing for their evolution with time (with redshift used as a proxy). Due to the high-dimensional nature of the required analysis, confronting the different models using likelihood-based pipelines has been cumbersome and reliant mainly on visual inspection of the results of parameter inference, e.g. comparing the dust-law distributions of high- and low-mass galaxies or examining trends of inferred properties with redshift. SBI, however, provides an avenue towards principled Bayesian model comparison by giving direct access to marginalized model probabilities (and Bayes factors) even for models with thousands of dimensions like the one in this study, as recently demonstrated by Karchev et al. (2023a). Furthermore, amortization allows for exploring the results as a function of the values of the underlying parameters on mock data – unthinkable with likelihood-based methods.

Selection effects and non-Ia contamination thus remain the two major hurdles left before a fully-fledged application of SBI to SN Ia cosmology becomes viable. Accounting for them requires simulating the way transients are identified and classified into a survey release, leading to different-sized surveys in the training set and transients observed in different time/band configurations. One might imagine two ways to circumvent these problems. One is to take a step back in the data processing pipeline, simulating and subsequently feeding in the neural ratio estimator raw telescope images, which have a set number, order, and characteristics after the survey has been performed (see, e.g. Sánchez et al. (2022) for a similar approach but using the conventional SN cosmology pipeline). This is equivalent to padding the light curves, which would be the standard approach to unequal-length sequences in machine learning, and including light curves for undetected objects. The downside is clear: the network must learn selection effects and contamination from a prohibitively large amount of mostly uninformative data. Alternatively, one might still try to condition the simulator on the observed light curves of detected objects by smartly modifying the hierarchical priors. However, this might cause issues when many objects are considered, as we discuss in Karchev et al. (in preparation).

In light of this, our current approach – conditioning on the number and order of SNæ in a survey and on the number and order of observations of each SN – has limited prospects of solving selection effects. In upcoming work, we will demonstrate cosmological inference in the presence of selection effects and variable-length data by using tools that have already been exploited in the SN literature: e.g. Gaussian process regression for regularizing the light curves (e.g. Revsbech et al. 2018; Boone 2019), in combination with cutting-edge techniques for permutation-invariant SBI (Rodrigues et al. 2021; Campeau-Poirier et al. 2023; Heinrich et al. 2024; Makinen, Alsing & Wandelt 2023).

We make use of `Clipppy`,[16] a Python package based on `pyro` (Bingham et al. 2019) and `PyTorch` (Paszke et al. 2019), for the probabilistic part of our forward simulator and `PyTorch Lightning` (Falcon & The PyTorch Lightning team 2023) for training the inference network.

---

[16] https://github.com/kosiokarchev/clipppy

## ACKNOWLEDGEMENTS

RT acknowledges co-funding from Next Generation EU, in the context of the National Recovery and Resilience Plan, Investment PE1 - 'Project FAIR Future Artificial Intelligence Research'. This resource was co-financed by the Next Generation EU [DM 1555 del 11.10.22]. RT is partially supported by the Fondazione ICSC, Spoke 3 'Astrophysics and Cosmos Observations', Piano Nazionale di Ripresa e Resilienza Project ID CN00000013 'Italian Research Center on High-Performance Computing, Big Data and Quantum Computing' funded by MUR Missione 4 Componente 2 Investimento 1.4: Potenziamento strutture di ricerca e creazione di 'campioni nazionali di R&S (M4C2-19)' - Next Generation EU (NGEU).

MG and KSM are supported by the European Union's Horizon 2020 research and innovation programme under ERC Grant Agreement No. 101002652 and Marie Sklodowska-Curie Grant Agreement No. 873089. BMB is supported by the Cambridge Centre for Doctoral Training in Data-Intensive Science funded by the UK Science and Technology Facilities Council (STFC).

CW has received funding from the European Research Council (ERC) under the European Union's Horizon 2020 research and innovation programme (Grant agreement No. 864035).

Part of this work was performed using resources provided by the Cambridge Service for Data Driven Discovery (CSD3) operated by the University of Cambridge Research Computing Service (www.csd3.cam.ac.uk), provided by Dell EMC and Intel using Tier-2 funding from the Engineering and Physical Sciences Research Council (capital grant EP/T022159/1), and DiRAC funding from the Science and Technology Facilities Council (www.dirac.ac.uk).

## DATA AVAILABILITY

This article makes use of data released with SNANA, available on Zenodo at https://dx.doi.org/10.5281/zenodo.4001177. The mock data generated in this research will be shared upon reasonable request to the corresponding author.

## APPENDIX A: BAYESIAN VALIDATION AND FREQUENTIST CALIBRATION

Amortized inference allows validating the coverage properties of the approximate posteriors (in a Bayesian sense) and producing confidence regions with exact frequentist coverage, as we detailed in SICRET Subsection 3.4. Here, we briefly explain the two procedures and related concepts and implement them, respectively, for Figs A1 and A2.

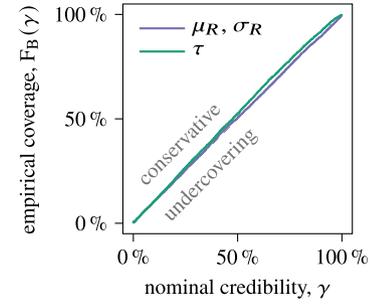

**Figure A1.** Bayesian P–P plot for global-parameter approximate posteriors trained for inference from the real data set.

We consider approximate posteriors $q(\boldsymbol{\Theta}|\boldsymbol{d}) \equiv \hat{r}(\boldsymbol{\Theta}, \boldsymbol{d})\,p(\boldsymbol{\Theta})$ evaluated for data $\boldsymbol{d}$ simulated from true parameter values $\boldsymbol{\Theta}_0$ and the corresponding credibilities

$$\gamma(\boldsymbol{\Theta}_0, \boldsymbol{d}) \equiv \int_{\Gamma_{\boldsymbol{\Theta}}(\boldsymbol{\Theta}_0, \boldsymbol{d})} q(\boldsymbol{\Theta}|\boldsymbol{d})\,\mathrm{d}\boldsymbol{\Theta} \tag{A1}$$

of highest-likelihood regions (HLRs)[17] $\Gamma_{\boldsymbol{\Theta}}(\boldsymbol{\Theta}_0, \boldsymbol{d}) \equiv \{\boldsymbol{\Theta} : \hat{r}(\boldsymbol{\Theta}\boldsymbol{d}) > \hat{r}(\boldsymbol{\Theta}_0, \boldsymbol{d})\}$.

Bayesian validation uses random parameters sampled from the prior: $\boldsymbol{\Theta}_0 \sim p(\boldsymbol{\Theta}_0)$, and the self-consistency of the prior and data-averaged posterior – the fact that, when averaged over data sets sampled from the marginal likelihood $p(\boldsymbol{d}) \equiv \int p(\boldsymbol{d}|\boldsymbol{\Theta})\,p(\boldsymbol{\Theta})\,\mathrm{d}\boldsymbol{\Theta}$, the exact posterior $p(\boldsymbol{\Theta}|\boldsymbol{d})$ reverts to the prior $p(\boldsymbol{\Theta})$:

$$\mathbb{E}_{p(\boldsymbol{d})}[p(\boldsymbol{\Theta}|\boldsymbol{d})] \equiv \int p(\boldsymbol{\Theta}|\boldsymbol{d})p(\boldsymbol{d})\,\mathrm{d}\boldsymbol{d} = \int p(\boldsymbol{\Theta}, \boldsymbol{d})\,\mathrm{d}\boldsymbol{d} = p(\boldsymbol{\Theta}), \tag{A2}$$

and so credibilities computed with $q(\boldsymbol{\Theta}|\boldsymbol{d}) \to p(\boldsymbol{\Theta}|\boldsymbol{d})$ are uniformly distributed over [0; 1], or equivalently, have a cumulative distribution $F_B(\gamma) = \gamma$. On a probability–probability (P–P) plot, depicting $F$ versus $\gamma$, this manifests as a diagonal line. If, empirically, $F_B(\gamma) > \gamma$, the posteriors $q(\boldsymbol{\Theta}|\boldsymbol{d})$ are, on average, conservative: i.e. they cover true values more often than expected. And conversely, $F_B(\gamma) < \gamma$ implies they are overconfident, i.e. exhibit a greater scatter around the true value (or, possibly, even a bias) than expected from their sizes. However, $p(\boldsymbol{\Theta}|\boldsymbol{d})$ is not the only distribution which has perfect Bayesian coverage: in fact, using even the prior for $q(\boldsymbol{\Theta}|\boldsymbol{d})$ would lead to $F_B(\gamma) = \gamma$.

Instead of averaging over the prior, one can examine the distribution of credibilities, conditioning on a fixed parameter value $\boldsymbol{\Theta}_0$, which leads to $\boldsymbol{d} \sim p(\boldsymbol{d}|\boldsymbol{\Theta}_0)$ (instead of $\boldsymbol{d} \sim p(\boldsymbol{d})$) and so can be used for frequentist inference. In this scenario, in general, $F_f(\gamma|\boldsymbol{\Theta}_0) \neq \gamma$ due to the approximate nature of $q(\boldsymbol{\Theta}|\boldsymbol{d})$ on one hand, as before, but also because of the influence of a non-uniform prior. Calculating $F_f(\gamma|\boldsymbol{\Theta}_0)$ as a function of $\boldsymbol{\Theta}_0$, e.g. on a grid or by using nearest neighbours among prior samples (we use the latter), allows one to derive the *required credibility* $\hat{\gamma}(\boldsymbol{\Theta}_0\tilde{\gamma})$, for any desired confidence $\tilde{\gamma}$, as the $\tilde{\gamma}^\mathrm{th}$ quantile of $F_f(\gamma|\boldsymbol{\Theta}_0)$: see Karchev et al. (2023a, fig. 3) for an illustration.

---

[17]In SICRET, we instead used highest posterior density (HPD) regions. The difference is usually imperceptible, especially in well-constrained scenarios, and in fact any region definition can be used: here we use the approximate likelihood since it is directly accessible through $\hat{r}$ and independent of parametrization, while posterior densities require evaluating the prior density as well (or density estimation from weighted samples). Of note is also an alternative method of defining credible regions using distances to random points (DRP), as described by Lemos et al. (2023), which can, in certain circumstances, detect a systematic bias in $q(\boldsymbol{\Theta}|\boldsymbol{d})$.

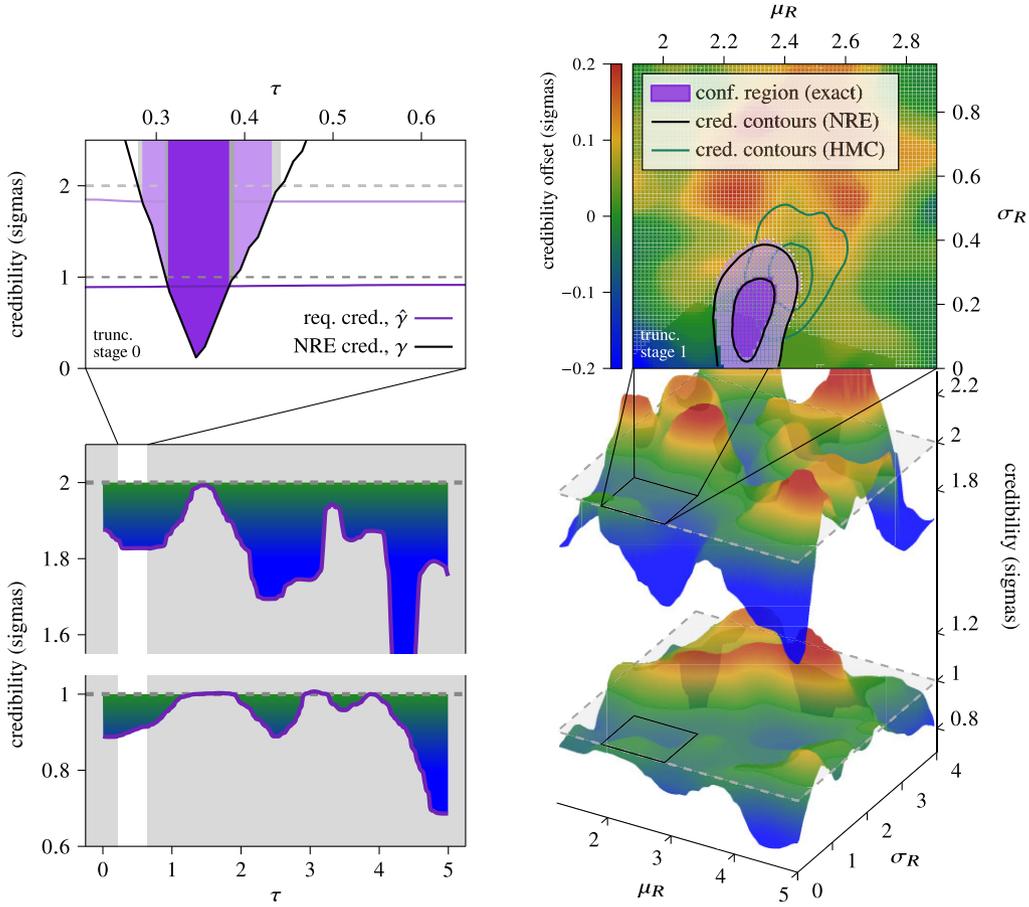

**Figure A2.** Calibrated frequentist global-parameter inference from the real CSP data set. As purple lines and coloured surfaces, we show the required credibility $\hat{\gamma}$ that achieves 1- or $2\sigma$ coverage (corresponding to 68.3 and 95.4 per cent in one dimension and $\approx 39$ and $\approx 86$ per cent in two dimensions), as a function of the parameter value. Everywhere, the colour axis represents the difference between the required credibility and the confidence level (empirical coverage). In the top row, using black lines, we depict nominal credibility from the NRE posterior evaluated on the observed data, which is used to derive calibrated 1- and $2\sigma$ confidence regions, filled in purple. The difference with the approximate posterior is insignificant, but the confidence region thus constructed has guaranteed coverage.

This paper has been typeset from a T<sub>E</sub>X/L<sup>A</sup>T<sub>E</sub>X file prepared by the author.